# A Novel Robust Fault Detection Scheme of Lipschitz Nonlinear Systems Using Combination of Bond Graph and Observer

Mohammad Ghasem Kazemi[1] and Mohsen Montazeri[2]

## Abstract

This paper deals with a new robust fault detection (FD) scheme for nonlinear Lipschitz systems wherein a robust nonlinear observer is used in combination with the Bond Graph (BG) method. In order to improve the efficiency of the classical Analytical Redundancy Relations (ARRs) FD scheme based on the BG method, a new form of the ARRs for the nonlinear Lipschitz systems is presented. This new form of the residuals is based on the output estimation error of the observer and is called Error-based Analytical Redundancy Relations (EARRs). The robustness against disturbances and parametric uncertainties in the FD system is achieved by the proposed method. Next, the integral form of the EARRs for the nonlinear Lipschitz system is presented that is robust in the sense of measurement noises as well. The BG model of a single link manipulator with revolute joints actuated by a DC motor is derived and the efficiency of the proposed method in comparison to the classical ARR method is shown in the simulation results.

## 1- Introduction

Fault diagnosis has become an important and vital issue in industrial applications. Different approaches including model based and data driven methods are presented in the literature for fault diagnosis in different dynamic systems. It is recognized that the model-based FD methods are more powerful than data-driven methods [1, 2] because it uses more information about the process. Model based methods are based on a mathematical model of the system and may be given greater efficiency provided that the exact model of the system is available. In other words, the mathematical model of the system plays the main role in the model based FD system design.

[1] *Department of Electrical and Computer Engineering IAU of Gachsaran, Iran.*

[1,2] *Department of Electrical and Computer Engineering, Shahid Beheshti University, A.C., Tehran, Iran.*

**Corresponding Author:**

Mohammad G. Kazemi, Department of Electrical and Computer Engineering Shahid Beheshti University, A.C., Tehran, Iran. orcid.org/0000-0002-0236-9974

Tel.: +989177428946

Fax: +98-21-77310425

Email: mg_kazemi@sbu.ac.ir

Bond graph method is introduced to represent the dynamic systems in various physical domains such as electrical, mechanical, hydraulic and etc. by a common language and may be considered in the model based FD system category. The state space representation and ARRs of a system may be derived from the BG model in some systematic approaches [3]. Several applications of the BG method for modeling and fault diagnosis of different linear systems, nonlinear systems, hybrid systems and etc. can be found in the literature. Some industrial processes such as gas turbine [4], mechanical phenomenon [5], wind turbine [6], boiler [7,8] and robots [9-11] are modeled based on the BG method. The main benefits of the BG method may be stated as the common language for different domains of energy, possibility of structural analysis, suitable sensor placement and appropriate fault detection, isolation and identification.

The BG method is based on the mass and energy conservation law. The defined elements in the model are based on physical concepts in different domains of energy and thus, it provides efficient fault diagnosis capabilities. In the BG method, the quantitative model based fault diagnosis scheme is based on the ARRs, which may be derived in some systematic approaches. The ARRs form the residuals of the FD system and Fault Signature Matrix (FSM) may be derived analytically based on the given ARRs, which will be used further for fault isolation and identification.

The robustness of the FD system is an important aspect that is also noticed in the BG method. Several studies on the robustness for the BG method are given in the literature. Two approaches for uncertainties modeling using bond graph method is proposed in [12], which further used for robust fault diagnosis purposes in different studies. The modeling and robust fault diagnosis of a DC motor based on the BG method is presented in [13] wherein the parameters of the BG model are derived by applying identification techniques to real data of the system. The parameter uncertainties are represented in Linear Fractional Transformation (LFT) form which further produces some adaptive thresholds on the residuals. In [8], the modeling and robust fault diagnosis of a steam generator using uncertain bond graph model is addressed. The BG method is used for different purposes such as nonlinear modeling of the considered steam generator; structural analysis, adaptive thresholds on the residuals and sensitivity analysis of the residuals. In [14], a robust bond graph model-based fault detection and isolation approach is developed in order to improve the robustness in the presence of measurements and parameters uncertainties. A robust fault diagnostic scheme for locomotive electro-pneumatic brake as a multi-energy domain system is presented in [15] that integrates a model-based strategy and a data-driven approach. The developed scheme is based on the LFT form uncertainties in the physical parameters of the BG modeling.

The robustness issue in the abovementioned studies is presented for parametric uncertainties in which the upper and lower limits on the BG model parameters are derived that define the upper and lower limits of the thresholds. This approach is called adaptive threshold in the literature [8] due to their variations according to the variations of the inputs and outputs of the system.

The robustness against the noise effects is tackled by the integral form of the residuals in [16], which is based on the Laplace transform of the residuals. It is clear that this form of the residuals cannot be used for nonlinear systems. Different criteria of the integral form of the residuals for incipient fault detection in linear system are studied in [17]. Another aspect of the robustness in the FD system is in the sense of disturbances in the system, which may have more effects on the residuals in comparison to noises or parametric uncertainties. One of the limited studies on the robustness against disturbances in the BG method is presented recently in [18-19], in which the proposed method may only be used for linear systems.

To the best of our knowledge, the problem of robust FD scheme for nonlinear systems based on the BG method considering the simultaneous effects of disturbances, noises and parametric uncertainties is not investigated in the literature. The aim of this paper is to design a robust FD system for the nonlinear Lipchitz systems in the sense of disturbances and noises, which also may enhance the robustness against parametric uncertainties. The robustness aspect against disturbance and parametric uncertainties are achieved by using a nonlinear Lipschitz observer in the FD system wherein the effects of disturbances are attenuated in the output estimation error and the Lipschitz constant is maximized simultaneously. The ARRs of the FD system are presented in a new form based on the output estimation error of the observer and the parameters of the BG model, which leads to more robustness of the FD system. It is assumed that the understudy system is in the Lipschitz nonlinear system class and therefore, both obtained ARRs and the state space equations of the BG model will satisfy the Lipschitz property. The observer design for the Lipschitz nonlinear systems is studied based on different approaches such as H∞, H2, H-/H∞ and so on, which will further be used for FD purposes [20-24], to list a few. In this paper, a new robust observer is presented based on the projection lemma, which simultaneously maximizes the Lipschitz constant and minimizes the disturbance attenuation level.

Succinctly, the main contributions of the paper are as follows.
- Presenting a new form of the ARRs for the nonlinear Lipschitz systems based on the output estimation error.
- Robust FD of the nonlinear Lipschitz systems based on the BG method with simultaneous robustness against disturbances, noises and parametric uncertainties.

The rest of this paper is considered as follows. In section 2, the problem formulation and preliminaries are presented. The proposed robust fault detection including the EARRs, robust observer, robustness against measurement noises and robustness against parametric uncertainties are given in section 3. The BG modeling and simulation results of the proposed FD scheme for a single-link manipulator with revolute joints actuated by a DC motor are provided as a Lipschitz nonlinear system in section 4, follows by conclusion in section 5.

## 2- Problem formulation and preliminaries

In this paper, the proposed FD system is designed based on the BG model. The BG model of a system may be provided or be derived based on the mass and energy conservation law. Different physical phenomena are modeled by defined elements in the BG method. The BG model of a dynamic system may be contained the resistive elements (R), capacitive elements (C), inductive elements (I), transformers (TF), gyrator (GY), effort sources (Se), flow sources (Sf), effort sensors (De), flow sensors (Df), 0-junctions and 1-junctions. The ARRs as the residuals of the FD system can be derived using the derivative causality BG model of the system. The generic form of the ARRs may be given as:

$$ARR = \psi(y^{(n)}, \dots, y, u^{(j)}, \dots, u, \theta) \quad (1)$$

where $y \in R^l$ is the output vector including taken information from the effort and flow sensors, $u \in R^m$ is the input vector including taken information from the effort and flow sources and $\theta$ is the parameters of the BG model (resistive, inductive and capacitive elements and the ratio of the gyrators and transformers elements). $n$ and $j$ are also denoted to the order of the outputs and inputs derivatives in an ARR, respectively. The number of the outputs and inputs are given as $l$ and m, respectively.

Generally, the ARRs are given according to known information of the system including the inputs, outputs and their derivatives and the given parameters in the BG model. In optimal manner, the number of the ARRs is equal to the number of sensors in the system. These ARRs may be used for fault detection, fault isolation and fault identification as well, which are obtained in nonlinear form for the nonlinear systems. The ARRs are mainly some differential equations that their orders are dependent on the sensor placement in the system. The ARRs with zeroth, first and second orders are useful for fault diagnosis purpose.

The residuals of the FD system are directly evaluated in according to the taken information of the inputs, outputs and the known or identified parameters of the BG model. The exceeded residuals indicate the fault occurrence, which further can be used for fault isolation and identification purposes. The goal of this study is to improve the efficiency of the FD system based on the BG method for a special class of the nonlinear systems as the nonlinear Lipschitz class. Thus, the problem definition is to derive a new form of the residuals that is robust against parametric uncertainties, disturbances and noises and will maintain the fault sensitivity in the FD system. These criteria may be satisfied by aid of the proposed combined BG-observer FD system, which is shown in Fig.1.

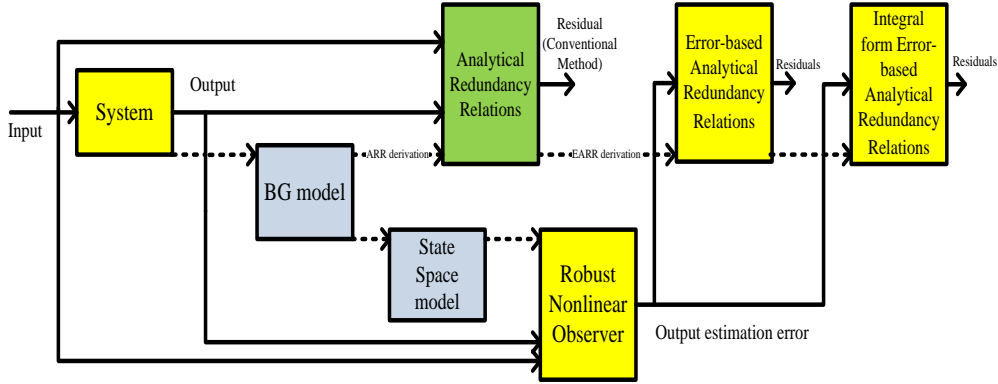

Figure 1. The procedure and structure of the proposed FD system

In the first stage, a nonlinear Lipschitz observer is designed based on the state space representation of the BG model, which attenuates the effects of disturbances on the output estimation error and is robust against parametric uncertainties. Next, these output estimation errors are given to the EARRs and Integral-form Error-based Analytical Redundancy Relations (IEARRs) of the BG model, which used as the residuals of the FD system. The robustness against noises may be achieved by the presented IEARRs method.

In the sequel of this section, some lemmas are given that are used in the main results.

**Lemma 1** [25]: For a given symmetric matrix $Z \in S_m$ and two matrices $U$ and $V$ with $m$ column dimension, there exists an unstructured matrix $X$ that satisfies the following inequality.

$$U^T X V + V^T X^T U + Z < 0 \qquad (2)$$

If and only if the inequalities (3) and (4) are satisfied with respect to decision matrix $X$.

$$N_U^T Z N_U < 0 \qquad (3)$$

$$N_V^T Z N_V < 0 \qquad (4)$$

where the columns of the arbitrary matrices $N_V$ and $N_U$ are the basis for null spaces of $U$ and $V$, respectively.

**Lemma 2** [26]: Let $D$, $S$ and $F$ are real matrices with proper dimensions. The unknown but bounded matrix $F$ satisfies the following condition.

$$F^T F \leq I \qquad (5)$$

Then, for any scalar $\epsilon > 0$ and vectors $x, y \in R^n$, the following inequality can be given:

$$2x^TDFSy \leq \frac{1}{\epsilon}x^TDD^Tx + \epsilon y^TS^TSy \qquad (6)$$

## 3- Main Results

The main results of this paper are given in four subsections including Error based Analytical redundancy relations, robust nonlinear observer for nonlinear Lipschitz system, robustness of the FD system against measurement noise by integral form of the residual and robustness against parametric uncertainties as follows.

### 3.1 Error based analytical redundancy relations

In order to achieve a robust FD system against disturbances and noises, a new form of ARRs is presented in this section. The ARRs may be in the static or dynamic form. The static form of the ARRs indicates the hardware redundancy in the system, which is a common approach in industrial applications. The dynamic ARRs are some differential equations, which may have different orders on the basis of the sensors location in the system. The following general form of the ARRs with the maximum second order assumption is considered in this paper.

$$ARR = \psi(\ddot{y}, \dot{y}, y, \ddot{u}, \dot{u}, u, \theta) \qquad (7)$$

where $\psi$ is a function with linear and nonlinear terms of the outputs, inputs and the parameters of the BG model. In this paper, it is assumed that the nonlinear term is a Lipschitz function and is only dependent on the parameters, inputs and outputs of the system, not their derivatives. The derived nonlinear state space representation for this class of the ARRs will satisfy the Lipschitz condition. As mentioned in [27], this class of the nonlinear Lipschitz systems may be considered as a general class, at least locally, and hence it is not too restrictive assumption.
Therefore, the general form of the ARRs may be given as below.

$$ARRs = M_1(\theta)\ddot{y}(t) + M_2(\theta)\dot{y}(t) + M_3(\theta)y(t) + N_1(\theta)\ddot{u}(t) + N_2(\theta)\dot{u}(t) + N_3(\theta)u(t) + \Psi(y, u, \theta) \qquad (8)$$

where $M_i$ and $N_i$ i=1,2,3 are matrices with appropriate dimensions based on the parameters of the BG model and $\Psi$ is the nonlinear term that satisfies the Lipschitz inequality in (9) with Lipschitz constant matrix $\Upsilon$.

$$\Psi(y, u, \theta) - \Psi(\hat{y}, u, \theta) \leq \Upsilon|y(t) - \hat{y}(t)| \qquad (9)$$

In the case that there are no faults, disturbances and noises in the system, the ARRs are zero provided that the given model has high degree of precision. If an observer is designed properly, the following estimated ARRs may also be given, which is zero in the normal case same as the ARRs.

$$\widehat{ARR} = M_1(\theta)\ddot{\hat{y}}(t) + M_2(\theta)\dot{\hat{y}}(t) + M_3(\theta)\hat{y}(t) + N_1(\theta)\ddot{u}(t) + N_2(\theta)\dot{u}(t) + N_3(\theta)u(t) + \Psi(\hat{y}, u, \theta) \qquad (10)$$

The difference between an ARR and its estimation is as (11).

$$\begin{aligned}ARR - \widehat{ARR} = & M_1(\theta)\ddot{y}(t) + M_2(\theta)\dot{y}(t) + M_3(\theta)y(t) + N_1(\theta)\ddot{u}(t) + N_2(\theta)\dot{u}(t) + N_3(\theta)u(t) + \Psi(y, u, \theta) \\ & - M_1(\theta)\ddot{\hat{y}}(t) - M_2(\theta)\dot{\hat{y}}(t) - M_3(\theta)\hat{y}(t) - N_1(\theta)\ddot{u}(t) - N_2(\theta)\dot{u}(t) - N_3(\theta)u(t) \\ & - \Psi(\hat{y}, u, \theta)\end{aligned} \qquad (11)$$

Equation (12) is achieved by simplifying (11), which is called EARR.

$$EARR = M_1(\theta)(\ddot{y}(t) - \ddot{\hat{y}}(t)) + M_2(\theta)(\dot{y}(t) - \dot{\hat{y}}(t)) + M_3(\theta)(y(t) - \hat{y}(t)) + \Psi(y, u, \theta) - \Psi(\hat{y}, u, \theta) \qquad (12)$$

As it is clear from (12), the direct effects of the inputs on the residuals are removed in the linear terms of the new defined form. It is also worth noting that the nonlinear terms according to the inputs only will be removed in the EARRs, which leads to simpler residuals. The effects of the removed inputs are remained in the residuals in indirect manner using the observer. The derivatives of the inputs in the ARRs may lead to false alarm due to their abrupt changes, which is obviated in the proposed method.

As we know

$$e_y(t) = y(t) - \hat{y}(t) \tag{13}$$

The following equations are obtained by taking derivative of (13).

$$\dot{e}_y(t) = \dot{y}(t) - \dot{\hat{y}}(t) \tag{14}$$

$$\ddot{e}_y(t) = \ddot{y}(t) - \ddot{\hat{y}}(t) \tag{15}$$

It is worth noting that the given derivatives of the output estimation errors have better performance in comparison to the derivatives of the inputs and outputs of the system. This is because of this fact that the adverse effects of disturbances and abrupt changes in the outputs and inputs are reduced by the observer in the proposed method. The output estimation error of the observer is achievable from the observer and their derivative with reduced adverse effects may be calculated in the FD system. On the other hand, the state space representation of the integral causality BG model may be easily derived which can be used for the observer design.

Considering (13), (14) and (15), the EARR may be given as follows.

$$EARR = M_1(\theta)\ddot{e}_y(t) + M_2(\theta)\dot{e}_y(t) + M_3(\theta)e_y(t) + \Psi(y, u, \theta) - \Psi(\hat{y}, u, \theta) \tag{16}$$

According to the Lipschitz inequality, the upper limit of the EARR may be given in (17).

$$EARR = M_1(\theta)\ddot{e}_y(t) + M_2(\theta)\dot{e}_y(t) + M_3(\theta)e_y(t) + \Upsilon e_y(t) \tag{17}$$

Equation (17) can be expressed in linear form by defining a new matrix as $M_{3n}$ as follows:

$$EARR = M_1(\theta)\ddot{e}_y(t) + M_2(\theta)\dot{e}_y(t) + M_{3n}(\theta, \gamma)e_y(t) \tag{18}$$

where

$$M_{3n}(\theta, \gamma) = M_3(\theta) + \Upsilon I \tag{19}$$

In fact, a new linear form of the residuals are obtained using the Lipschitz inequality and considering the upper limit of the Lipschitz term, which may provide the facility to define the integral form of the ARRs for the nonlinear Lipschitz systems. The new defined form is given according to the output estimation error of the observer and will also be zero in the normal case provided that the observer is designed properly.

Same as the conventional ARRs, the obtained residual is also sensitive to the effects of disturbances and faults in the system, but there is the possibility of some trade-off between these effects. In other words, the possibility of disturbance attenuation is well provided in the proposed method, which may not be achieved in the conventional ARRs. Generally, a combined BG-observer based FD system is proposed wherein different criteria of an FD system are defined in each stage. The robustness against disturbance and parametric uncertainties are defined in the FD system based on the observer. The fault sensitivity, robustness against noises and possibility of FSM derivation, and possibility of fault detection, isolation and identification are also provided in the BG part based on the new defined residuals named as EARRs and IEARRs. The utilized robust observer for the nonlinear Lipschitz system is presented in the following subsection.

## 3.2 Robust Nonlinear Observer for nonlinear Lipschitz system

Because of the defined residuals, which are based on the output estimation error, the observer must be defined in such a way that the effects of disturbances are removed or attenuated in the output estimation error, while maintain the fault sensitivity. In this paper, a new observer for the nonlinear Lipschitz system is presented based on the projection lemma, which may reduce the conservatism of the FD system [28]. The Lipschitz constant is also assumed as one of the optimization variables, which must be maximized in the observer design.

The following state space representation for the nonlinear Lipschitz system may be achieved by using the defined states in the BG method. These states are the integral of effort variables for the inductive elements and the integral of flow variables for the capacitance elements in integral causality BG model of the system.

$$\dot{x}(t) = Ax(t) + Bu(t) + \varphi(x,u) + D_1 d(t) + Q_1 f(t) \tag{20}$$

$$y(t) = Cx(t) + D_2 d(t) + Q_2 f(t) \tag{21}$$

where $x \in R^n, u \in R^m, y \in R^l, d \in R^{k1}$ and $f \in R^{k2}$, which are denoted to the states, inputs, outputs, unknown exogenous disturbances and faults in the system, respectively. A, B, C, $D_1$, $Q_1$, $D_2$ and $Q_2$ are matrices with appropriate dimensions. The nonlinear term is defined as $\varphi(x,u)$ and is assumed Lipschitz, which satisfies:

$$\|\varphi(x,u) - \varphi(\hat{x},u)\| \leq \gamma \|x - \hat{x}\| \tag{22}$$

According to [26], any nonlinear system of the form $\dot{x} = f(x,u)$ can be expressed in the considered form (20)-(21), as long as $f(x,u)$ is differentiable with respect to $x$.

The Luenberger observer in (23) is used to estimate the states of the nonlinear Lipschitz system [27,29,30]

$$\dot{\hat{x}}(t) = A\hat{x}(t) + Bu(t) + \varphi(\hat{x}(t),u(t)) + L(y(t) - \hat{y}(t)) \tag{23}$$

$$\hat{y}(t) = C\hat{x}(t) \tag{24}$$

where $L$ is the gain of the observer.

The dynamic error of the observer may be achieved as (26) by defining the state estimation error in (25).

$$e(t) = x(t) - \hat{x}(t) \tag{25}$$

$$\dot{e}(t) = (A - LC)e(t) + \varphi(x,u) - \varphi(\hat{x},u) + (D_1 - LD_2)d(t) + (Q_1 - LQ_2)f(t) \tag{26}$$

The output estimation error of the observer may be given as (27).

$$e_y(t) = y(t) - \hat{y}(t) = Ce(t) + D_2 d(t) + Q_2 f(t) \tag{27}$$

The disturbance attenuation criterion of the observer is defined as an $H_\infty$ performance on the output estimation error as follows.

$$\frac{\|e_y(t)\|_2}{\|d(t)\|_2} \leq \mu \tag{28}$$

which may be given as

$$e_y(t)^T e_y(t) - \mu^2 d(t)^T d(t) \leq 0 \tag{29}$$

The following theorem is given for the robust nonlinear observer design based on the projection lemma.

**Theorem:** Consider the nonlinear Lipschitz system (20)-(21) along with the observer as (23). The state estimation error dynamics (26) are asymptotically stable with a predefined decay rate and the output estimation error of the observer satisfies the $H_\infty$ performance (28) for any 2-bounded norm disturbance signal and maximize the Lipschitz constant, if for given $\beta > 0$ and $\lambda > 0$, there exists the scalars values $\varepsilon > 0, \gamma > 0$ and the matrices $P > 0$, $X$ and $N$ such that the following LMI optimization problem has solution.

$$\min \quad \varepsilon - \gamma$$

$$\text{s.t}$$

$$\begin{bmatrix} C^T C + 2\beta P & \gamma I & P & P + \lambda A^T X - \lambda C^T N^T & C^T D_2 \\ * & -I & 0 & 0 & 0 \\ * & * & -I & 0 & 0 \\ * & * & * & -\lambda X - \lambda X^T & \lambda X^T D_1 - \lambda N D_2 \\ * & * & * & * & D_2^T D_2 - \varepsilon I \end{bmatrix} < 0 \quad (30)$$

The observer gain and the obtained disturbance attenuation level are given as:

$$L = X^{-T} N, \mu = \sqrt{\varepsilon}$$

**Proof.** Given the Lyapunov function as (31) its derivative may be obtained as (32).

$$V(t) = e(t)^T P e(t) \quad (31)$$

$$\dot{V} = \dot{e}(t)^T P e(t) + e(t)^T P \dot{e}(t) = e(t)^T A_{cl}^T P e(t) + e(t)^T P A_{cl} e(t) + 2e(t)^T P(\varphi - \hat{\varphi}) + 2e(t)^T P D_{cl} d(t) + 2e(t)^T P Q_{cl} f(t) \quad (32)$$

where

$$A_{cl} = A - LC, D_{cl} = D_1 - LD_2, Q_{cl} = Q_1 - LQ_2 \quad (33)$$

Based on Lemma 2, it may be given that:

$$2e(t)^T P(\varphi - \hat{\varphi}) \leq e(t)^T PP e(t) + \gamma^2 e(t)^T e(t) \quad (34)$$

which yield to

$$\dot{V}(t) \leq e(t)^T (A_{cl}^T P + PA_{cl} + 2\beta P + PP + \gamma^2 I) e(t) + 2e(t)^T P D_{cl} d(t) + 2e(t)^T P Q_{cl} f(t) \quad (35)$$

where $\beta$ is used to define the decay rate of the observer.
The upper bound of the Lyapunov function derivative in fault free case is given in (36).

$$\dot{V}(t) \leq e(t)^T (A_{cl}^T P + PA_{cl} + 2\beta P + PP + \gamma^2 I) e(t) + 2e(t)^T P D_{cl} d(t) \quad (36)$$

According to [31], we have:

$$e_y(t)^T e_y(t) - \mu^2 d(t)^T d(t) + \dot{V}(t) \leq 0 \quad (37)$$

Thus:

$$e(t)^T \left(C^TC + A_{cl}^T P + PA_{cl} + 2\beta P + PP + \gamma^2 I\right) e(t) + 2e(t)^T(PD_{cl} + C^T D_2)d(t) + d(t)^T(D_2^T D_2 - \mu^2 I)d(t) \tag{38}$$
$$\leq 0$$

Equation (38) can be written in the following LMI form

$$\begin{bmatrix} A_{cl}^T P + PA_{cl} + 2\beta P + PP + \gamma^2 I & PD_{cl} \\ * & 0 \end{bmatrix} + \begin{bmatrix} C^T C & C^T D_2 \\ * & D_2^T D_2 - \mu^2 I \end{bmatrix} \leq 0 \tag{39}$$

which can be given as (40)

$$\begin{bmatrix} I & A_{cl}^T \\ 0 & D_{cl}^T \end{bmatrix} \begin{bmatrix} 2\beta P + PP + \gamma^2 I & P \\ * & 0 \end{bmatrix} \begin{bmatrix} I & 0 \\ A_{cl} & D_{cl} \end{bmatrix} + \begin{bmatrix} 0 & C^T \\ I & D_2^T \end{bmatrix} \begin{bmatrix} -\mu^2 I & 0 \\ 0 & I \end{bmatrix} \begin{bmatrix} 0 & I \\ C & D_2 \end{bmatrix} < 0 \tag{40}$$

Thus, it can be concluded that:

$$\begin{bmatrix} I & A_{cl}^T & 0 \\ 0 & D_{cl}^T & I \end{bmatrix} \cdot \begin{bmatrix} 2\beta P + PP + \gamma^2 I + C^T C & P & C^T D_2 \\ * & 0 & 0 \\ * & * & D_2^T D_2 - \mu^2 I \end{bmatrix} \cdot \begin{bmatrix} I & 0 \\ A_{cl} & D_{cl} \\ 0 & I \end{bmatrix} < 0 \tag{41}$$

By assuming (41) as $N_U^T Z N_U < 0$, it can be calculated that

$$U = [A_{cl} \quad -I \quad D_{cl}]$$

$N_V$ and $V$ are also assumed as

$$N_V = \begin{bmatrix} \lambda I & 0 \\ 0 & 0 \\ 0 & I \end{bmatrix}, V = [0 \quad \lambda I \quad 0]$$

Then, the following nonlinear matrix inequality may be achieved as (2) in Lemma 1.

$$\begin{bmatrix} 2\beta P + PP + \gamma^2 I + C^T C & P + \lambda A_{cl}^T X & C^T D_2 \\ * & -\lambda X - \lambda X^T & \lambda X^T D_{cl} \\ * & * & D_2^T D_2 - \varepsilon I \end{bmatrix} < 0 \tag{42}$$

where $\varepsilon = \mu^2$.
Using the Schur complement and substituting (33) in (42), the given LMI (30) in the Theorem can be achieved and the proof is completed.

### 3.3 Robustness of the FD system against measurement noise by integral form of the residual

The integral form of the residuals for linear ARRs is presented in [16] in order to consider the effects of measurement noises on the residuals. The integral form of the ARRs is given in (43), which may be used for the linear form EARRs as well in (44).

$$\hat{r}_{ARR} = s^{-n} \frac{d^n}{ds^n} ARR(s) \tag{43}$$

$$\hat{r}_{EARR} = s^{-n} \frac{d^n}{ds^n} EARR(s) \tag{44}$$

The *EARRs(s)* in frequency domain are obtained by taking the Laplace transform. The initial conditions are removed in the obtained *EARRs(s)* by taking *n*-times derivative with respect to *s*. The order of the derivatives is defined by the order of the *EARRs*. Then, $s^{-n}$, which indicate *n*-times integration, is multiplied by the obtained derivative of the *EARR(s)*. Finally, the integral form of the EARRs is obtained by inverse Laplace transform, which is given as the IEARRs.

$$IEARR = \mathcal{L}^{-1}\{\hat{r}_{EARR}\} \tag{45}$$

A lemma is presented here to calculate the integral form of the residuals.
**Lemma 3:** Consider the following linear form of the EARRs.

$$EARR = M_1(\theta)\ddot{e}(t) + M_2(\theta)\dot{e}(t) + M_3(\theta)e(t) \tag{46}$$

The integral form of the abovementioned linear-form EARRs, which is called IEARRs, can be given as (47).

$$IEARR: \iint \left(2M_1 e(t) - 2M_2 t e(t) + M_3 t^2 e(t)\right) dt dt + \int \left(-4M_1 t e(t) + M_2 t^2 e(t)\right) dt + M_1 t^2 e(t) \tag{47}$$

**Proof:** The *EARR(s)* can be achieved as (48) by taking Laplace transform of (46).

$$EARR(s): M_1 s^2 E(s) - M_1 s e(0) + M_1 \dot{e}(0) + M_2 s E(s) - M_2 e(0) + M_3 E(s) \tag{48}$$

Then, by taking two times derivative of the *EARR(s)* against *s* variable, the initial conditions are removed in the *EARR(s)* as follows.

$$\frac{d(EARR(s))}{ds}: 2sM_1 E(s) + M_1 s^2 \frac{dE(s)}{ds} - M_1 e(0) + M_2 E(s) + M_2 s \frac{dE(s)}{ds} + M_3 \frac{dE(s)}{ds} \tag{49}$$

$$\frac{d^2(EARR(s))}{ds^2}: 2M_1 E(s) + 2sM_1 \frac{dE(s)}{ds} + 2sM_1 \frac{dE(s)}{ds} + M_1 s^2 \frac{d^2 E(s)}{ds^2} + M_2 \frac{dE(s)}{ds} + M_2 \frac{dE(s)}{ds} + M_2 s \frac{d^2 E(s)}{ds^2} + M_3 \frac{d^2 E(s)}{ds^2} \tag{50}$$

The order-sorted equation may be achieved as (51).

$$\frac{d^2(EARR(s))}{ds^2}: \left(2M_1 E(s) + 2M_2 \frac{dE(s)}{ds} + M_3 \frac{d^2 E(s)}{ds^2}\right) + \left(4M_1 s \frac{dE(s)}{ds} + M_2 s \frac{d^2 E(s)}{ds^2}\right) + M_1 s^2 \frac{d^2 E(s)}{ds^2} \tag{51}$$

Finally, by multiplying (51) by $s^{-2}$ and taking inverse Laplace transform, the integral form of the EARRs i.e. IEARRs as in (47) can be achieved, which complete the proof. □

The residual evaluation function of the proposed FD system is regarded as:

$$J_L(t) = \sqrt{\int_{t_0+i}^{t_0+i+L_w} r(t)^T r(t) dt} \tag{52}$$

where $t_0$ is the first sample and $L_w$ is the length of the moving window on the residuals. The index *i* is also considered for the moving fixed length window and is increased with the arrival of the new sample of the residual. The threshold is considered as (53) for the worst case disturbance in the fault free case.

$$J_{th} = \min_{d(t),u(t)\in l_2,f(t)=0} J_L(t) \qquad (53)$$

### 3.4 Robustness of the FD system against parametric uncertainties

The proposed nonlinear Lipschitz observer is designed by simultaneous maximization of the Lipschitz constant and minimization of the disturbance attenuation level. It can be shown that this observer is also made robust against the parametric uncertainties in the model as well as the nonlinear uncertainty, i.e. the uncertainty in the Lipschitz constant.

Consider the following state equation of the nonlinear Lipschitz systems.

$$\dot{x}(t) = (A + \Delta A)x(t) + (B + \Delta B)u(t) + \varphi_1(x,u) \qquad (54)$$

According to [32,33], the uncertain parts of (54) may be written as (55).

$$\Delta A = M_1 F N_1 \quad \Delta B = M_2 F N_2 \qquad (55)$$

in which $M_i$ and $N_i$ $i=1,2$ are constant matrices. The $F$ matrix is also has the bounded 2-norm as follows:

$$\|F\|_2 \leq I, \forall t \in [0,\infty) \qquad (56)$$

Hence, (54) can be given as

$$\dot{x}(t) = Ax(t) + Bu(t) + \varphi_2(x,u) \qquad (57)$$

where

$$\varphi_2(x,u) = \Delta A\, x(t) + \Delta B u(t) + \varphi_1(x,u) \qquad (58)$$

It is easily proved that the new defined nonlinear term is also a Lipschitz function, but with a larger Lipschitz constant, which can be obtained by maximization of the Lipschitz constant.

In the following section of the paper, the single link manipulator is provided to show the efficiency of the proposed FD scheme in comparison to the conventional ARRs method.

## 4- Simulation results

A single-link manipulator with revolute joints actuated by a DC motor as Fig.2 is given to study the performance of the proposed method in derivative (EARRs) and integral (IEARR) forms in comparison to the conventional method (ARRs). The observer design and fault detection of the considered manipulator are studied in several papers such as [34-36].

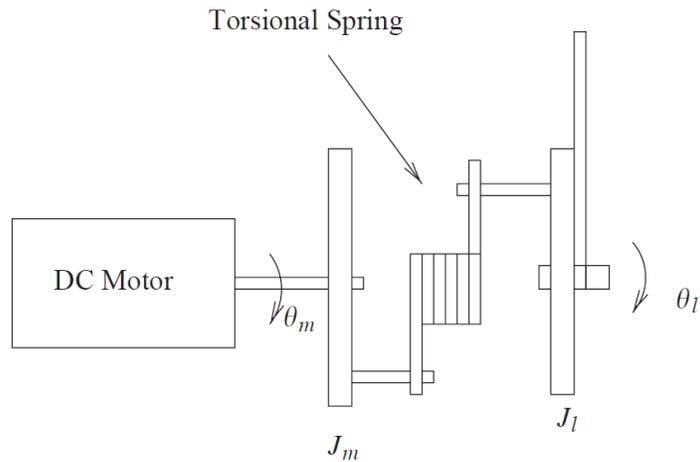

Figure 2. Schematic of the single-link manipulator with revolute joints actuated by a DC motor [37]

The parameters of the model and their values are given in Table 1 [36].

Table1: The parameters of the single-link manipulator and their values [30]

| Symbol | Definition | Value |
| --- | --- | --- |
| $J_l$ | Link inertia | 9.3e-3 kg.m$^2$ |
| $J_m$ | Motor inertia | 3.7e-3 kg.m$^2$ |
| K | Torsional spring constant | 1.8e-1 Nm/rad |
| m | Link mass | 2.1e-1 kg |
| h | Link length | 0.15 m |
| $k_\tau$ | Amplifier gain | 8e-2 Nm/V |
| $B_m$ | Viscous friction coefficient | 4.6e-2 Nm/V |
| g | Gravity constant | 9.8 m/s$^2$ |

The derived integral causality bond graph model of the understudy system is shown in Fig.3

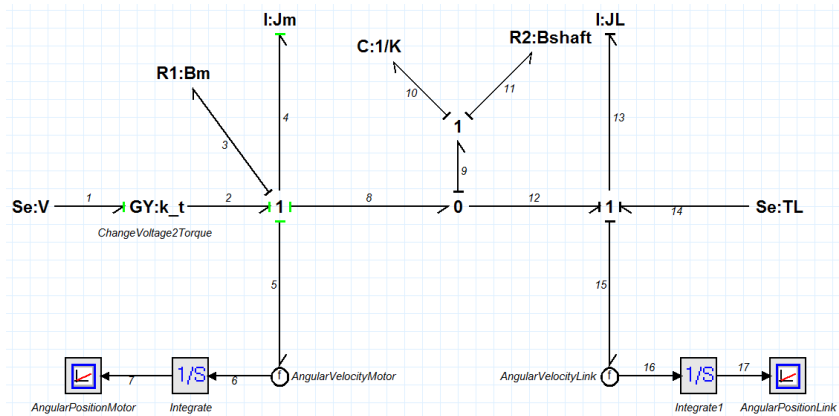

Fig3. BG model of single-link manipulator in integral causality in which, $B_m$ and $B_{shaft}$ are resistive elements that represent the viscous friction in the motor bearing [29] and damping of the shaft, respectively. The inductive elements of the BG model including $I:J_m$ and $I:J_l$ are respectively used for the inertia of the motor and link. The capacitance element $C=1/K$ defines the torsional spring of the shaft with spring constant $K$. The effort source $S_e:V$ is the DC voltage as the control input of the manipulator. The gyrator element $GY:k_t$ is used to define the relation between the applied voltage and the obtained torque of the motor by the constant of proportionality $k_t$. The $Se:T_L$ is the moment of gravitational force around the center of the link. Finally, the 0-junction and 1-junctions in the model are used to define common torques and common angular velocities in the model, respectively.

The simplified BG model of the manipulator is given in Fig.4 by neglecting the damping of the shaft.

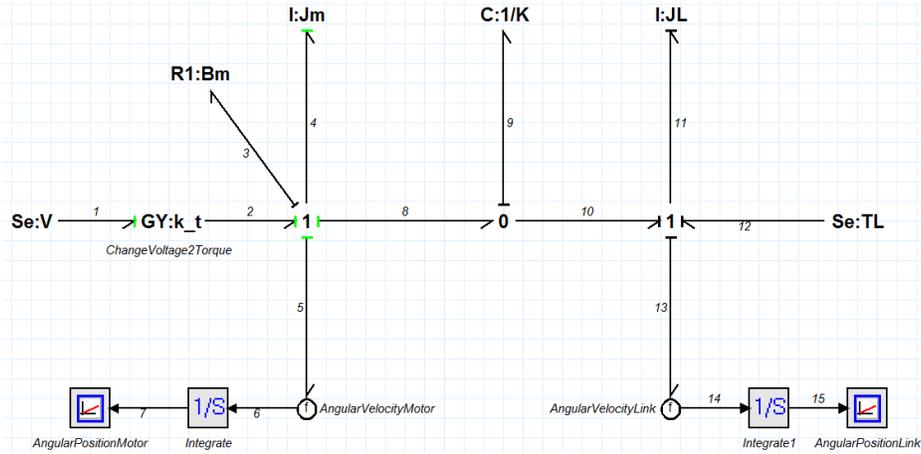

Fig4. Simplified BG model of single-link manipulator in integral causality

The ARRs of the system may be obtained using the Causality Inversion Method (CIM) [3] for the derivative causality BG model in Fig5.

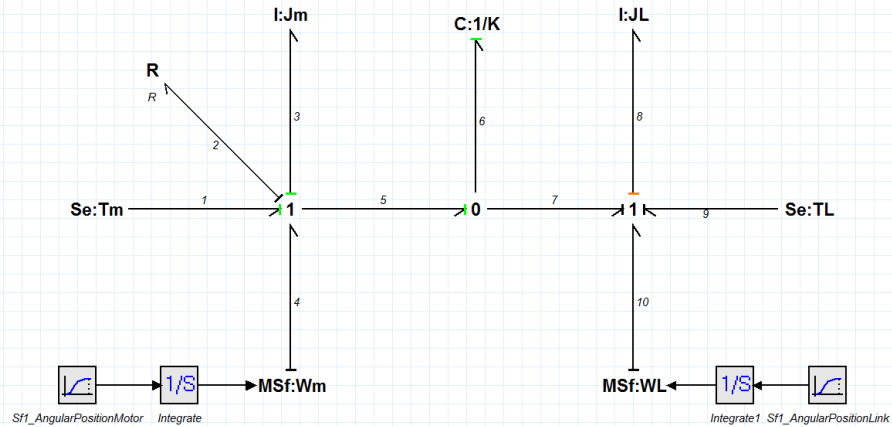

Fig5. BG model of single-link manipulator in derivative causality

$$\text{ARR1:} J_m \ddot{y}_1 + B\dot{y}_1 + K(y_1 - y_2) - k_\tau u \tag{59}$$

$$\text{ARR2:} J_l \ddot{y}_2 - K(y_1 - y_2) - mgh \sin y_2 \tag{60}$$

in which

$$y = [y_1 \quad y_2] = [\theta_m \quad \theta_l]$$

The considered matrices and the nonlinear function in (12) can easily be obtained as:

$$M_1 = \begin{bmatrix} J_m & 0 \\ 0 & J_l \end{bmatrix}, M_2 = \begin{bmatrix} B & 0 \\ 0 & 0 \end{bmatrix}, M_3 = \begin{bmatrix} K & -K \\ -K & K \end{bmatrix}$$

$$N_1 = 0, N_2 = 0, N_3 = \begin{bmatrix} -k_\tau \\ 0 \end{bmatrix}$$

$$\Psi(y, u, \theta) = \begin{bmatrix} 0 \\ -mgh \sin y_2 \end{bmatrix}$$

Thus, the EARRs are achieved as:

$$\text{EARR1:} J_m \ddot{e}_{y1} + B\dot{e}_{y1} + Ke_{y1} - Ke_{y2} \tag{61}$$

$$\text{EARR2:} J_l \ddot{e}_{y2} - Ke_{y1} + Ke_{y2} - mgh(\sin y_2 - \sin \hat{y}_2) \tag{62}$$

which can be given in linear form in (63)-(64).

$$\text{EARR1:} J_m \ddot{e}_{y1} + B\dot{e}_{y1} + Ke_{y1} - Ke_{y2} \tag{63}$$

$$\text{EARR2:} J_l \ddot{e}_{y2} - Ke_{y1} + Ke_{y2} - \gamma e_{y2} \tag{64}$$

Hence, the new defined matrix $M_{3n}$ in (19) may be written as.

$$M_{3n} = \begin{bmatrix} K & -K \\ -K & K - \gamma \end{bmatrix}$$

Moreover, the integral form of the residuals according to (47) and Lemma 3 are given in (65).

$$\text{IEARR:} \int\int\left(\begin{bmatrix} 2J_m & 0 \\ 0 & 2J_l \end{bmatrix} - \begin{bmatrix} 2Bt & 0 \\ 0 & 0 \end{bmatrix} + \begin{bmatrix} Kt^2 & -Kt^2 \\ -Kt^2 & Kt^2 - t^2\gamma \end{bmatrix}\right)\begin{bmatrix} e_{y1} \\ e_{y2} \end{bmatrix} dt\, dt \tag{65}$$
$$+ \int\left(\begin{bmatrix} -4J_m t & 0 \\ 0 & -4J_l t \end{bmatrix} + \begin{bmatrix} Bt^2 & 0 \\ 0 & 0 \end{bmatrix}\right)\begin{bmatrix} e_{y1} \\ e_{y2} \end{bmatrix} dt + \begin{bmatrix} J_m t^2 & 0 \\ 0 & J_l t^2 \end{bmatrix}\begin{bmatrix} e_{y1} \\ e_{y2} \end{bmatrix}$$

which can be written as follows.

$$\text{IEARR1:} \int\int (2J_m e_{y1} - 2Bt e_{y1} + Kt^2 e_{y1} - Kt^2 e_{y2})\, dt\, dt + \int(-4J_m t e_{y1} + Bt^2 e_{y1})\, dt + J_m t^2 e_{y1} \tag{66}$$

$$\text{IEARR2:} \int\int (2J_l e_{y2} - Kt^2 e_{y1} + Kt^2 e_{y2} - t^2\gamma e_{y2})\, dt\, dt + \int(-4J_l t e_{y2})\, dt + J_l t^2 e_{y2} \tag{67}$$

The state space representation of the system based on the integral causality BG model as Fig.4 may be achieved as follows, in which the sates are considered as the integral of effort variable for the inductive elements (p) and the integral of flow variable for the capacitive elements (q).

$$\dot{x}_1(t) = \frac{-B}{J_m} x_1(t) - Kx_2(t) + k_\tau u(t) \tag{68}$$

$$\dot{x}_2(t) = \frac{1}{J_m}x_1(t) - \frac{1}{J_l}x_3(t) \qquad (69)$$

$$\dot{x}_3(t) = Kx_2(t) - mgh\sin(x_4) \qquad (70)$$

$$\dot{x}_4(t) = \frac{1}{J_l}x_3(t) \qquad (71)$$

The physical meanings of the states are as follows.

$$x_1(t) = J_m\omega_m \qquad (72)$$

$$x_2(t) = \theta_{shaft} = \theta_m - \theta_l \qquad (73)$$

$$x_3(t) = J_l\omega_l \qquad (74)$$

$$x_4(t) = \theta_l \qquad (75)$$

in which $w_m$ is the angular velocity of the motor, $w_l$ is the angular velocity of the link, $\theta_m$ is the angular position of the motor and $\theta_l$ is the angular position of the link.

The obtained state space representation in (68)-(71) is same as the given one in [34-35] wherein the new states are considered. The relation between states may easily be given in (76).

$$z(t) = \begin{bmatrix} J_m\omega_m \\ \theta_m - \theta_l \\ J_l\omega_l \\ \theta_l \end{bmatrix} = Tx(t) = \begin{bmatrix} 0 & J_m & 0 & 0 \\ 1 & 0 & -1 & 0 \\ 0 & 0 & 0 & J_l \\ 0 & 0 & 1 & 0 \end{bmatrix}\begin{bmatrix} \theta_m \\ \omega_m \\ \theta_l \\ \omega_l \end{bmatrix} \qquad (76)$$

which leads to the following state space representation and validate the BG modeling of the system.

$$\dot{z}(t) = TAT^{-1}z(t) + TBu(t) + T\varphi(T^{-1}z, u) + TD_1d(t) + TQ_1f(t) \qquad (77)$$

$$y(t) = CT^{-1}z(t) + D_2d(t) + Q_2f(t) \qquad (78)$$

The disturbance and fault distribution matrix are also obtained using the given transformation from the model in [34]. Hence, the following matrices are achieved for the given state space representation of the system from the BG model.

$$A = \begin{bmatrix} -1.2432 & -0.1800 & 0 & 0 \\ 270.2703 & 0 & -270.2703 & 0 \\ 0 & 0.1800 & 0 & 1 \\ 0 & 0 & 107.5269 & 0 \end{bmatrix},$$

$$B = \begin{bmatrix} 0.08 \\ 0 \\ 0 \\ 0 \end{bmatrix}, C = \begin{bmatrix} 0 & 1 & 0 & 1 \\ 0 & 0 & 0 & 1 \end{bmatrix}$$

$$\varphi = \begin{bmatrix} 0 \\ 0 \\ -0.33\sin x_4 \\ 0 \end{bmatrix}, Q_1 = \begin{bmatrix} 0.08 & 1 \\ 0 & 0 \\ 0 & 0 \\ 0 & 0 \end{bmatrix},$$

$$D_1 = \begin{bmatrix} -0.0004 & 0.0001 & -0.0001 \\ -0.3000 & 0.0300 & -0.0600 \\ 0.0019 & 0.0002 & -0.0004 \\ 0.1000 & -0.0200 & 0.0400 \end{bmatrix},$$

$$D_2 = \begin{bmatrix} -0.001 & 0 & 0.003 \\ 0 & -0.010 & -0.002 \end{bmatrix}, Q_2 = \begin{bmatrix} 0 & 0.01 \\ 0.001 & 0 \end{bmatrix}$$

The disturbances on the considered single-link manipulator are assumed in sinusoidal form as follows [34].

$$d = [d_1 \quad d_2 \quad d_3]^T = [5\sin(10t) \quad 2\sin(10t) \quad \sin(20t)]^T \tag{79}$$

The observer gain, the disturbance attenuation level and the maximized Lipschitz constant are obtained as follows for $\beta = 1$ and $\lambda = 0.2$.

$$L = 1e3 \begin{bmatrix} 0.2141 & 0.1343 \\ 1.9888 & -0.6297 \\ 0.2487 & 0.1899 \\ 5.3438 & 4.0916 \end{bmatrix}, \mu = 0.2328, \gamma = 3.6529$$

The proposed FD system is simulated for different cases with and without measurement noises. At first, the simulation results are given for normal and faulty cases in the absence of measurement noises. Then, the performance of the proposed methods is compared with conventional ARRs method in the presence of measurement noises.

*Scenario 1: Fault free case in the presence of disturbances and in the absence of measurement noises*

In this scenario, the residuals are generated for the fault free case in the presence of disturbances. The thresholds on the residual evaluation function may be calculated for the worst case disturbance. The residuals and their evaluation functions for $t_0=80$ and $L_w=1000$ samples with 0.0001 sampling time are given in Fig.6 and Fig.7.

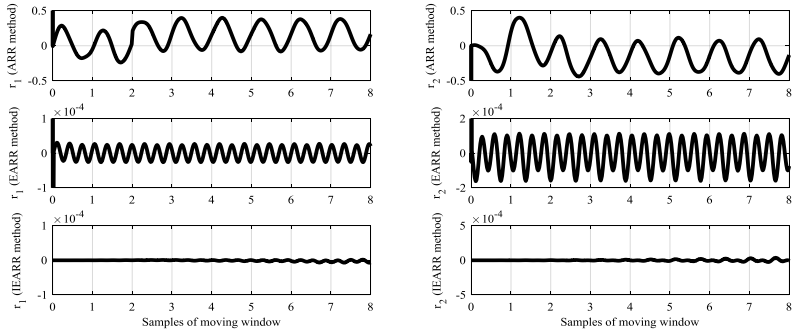

Fig 6: Residual for fault free case in the presence of disturbance

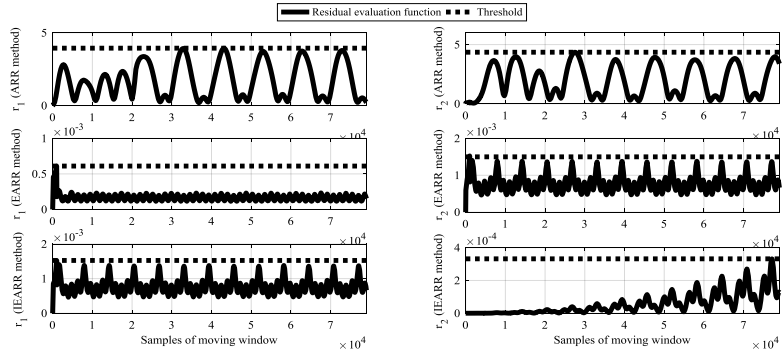

Fig.7: Residual evaluation function for fault free case in the presence of disturbance

The thresholds of the FD system may be obtained as Table 2.

Table 2. Threshold on the residual evaluation function

| Method | Residual $r_1$ | Residual $r_2$ |
|--------|----------------|----------------|
| ARRs   | 3.9419         | 4.3557         |
| EARRs  | 6.1028e-4      | 1.5012e-3      |
| IEARRs | 1.5252e-3      | 3.3191e-4      |

As it can be seen from the figures and the Table, the proposed methods including EARRs and IEARRs have better performance in fault free case in comparison to the conventional ARR method. The effects of disturbances are attenuated in the obtained residuals, while these effects lead to larger residuals in the conventional ARRs method.

*Scenario 2: Abrupt fault case in the presence of disturbances and in the absence of measurement noises*

In the second scenario, an abrupt fault is injected as the actuator fault. The injected fault is considered as a unit amplitude pulse that occurred at $t=6$ seconds. The residuals and their evaluation functions are shown in Fig.8 and Fig.9, respectively.

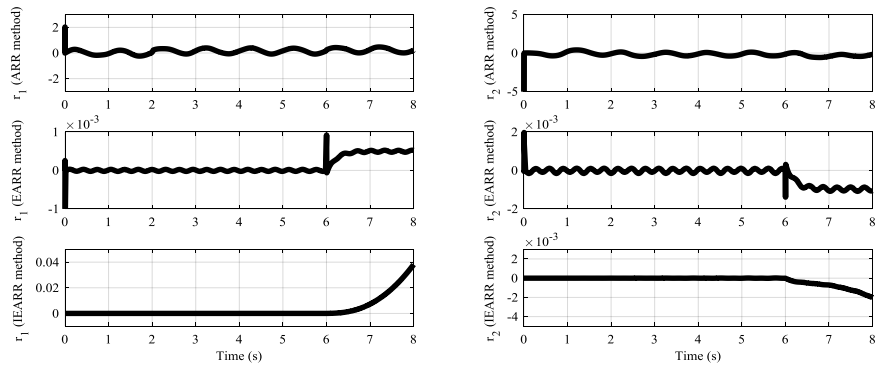

Fig 8: Residual for abrupt actuator fault

The residuals for the proposed methods show efficient changes due to occurred fault at $t=6$ seconds, while there is small variations in the residuals for the ARR method.

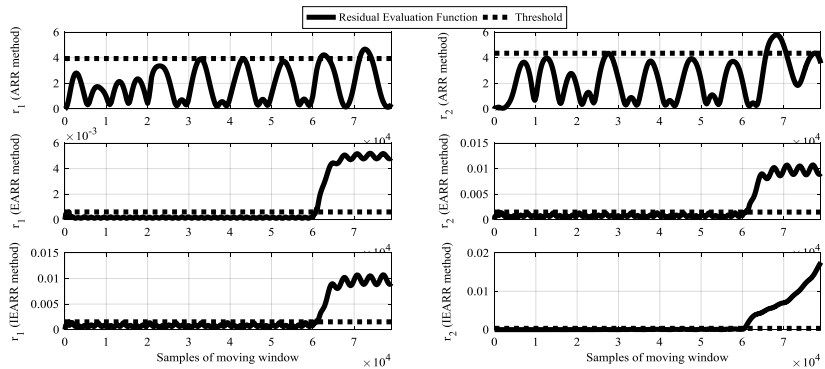

Fig.9 Residual evaluation function for abrupt actuator fault

The residual evaluation functions for the proposed FD system are exceeded from the defined thresholds in an efficient manner. The fault sensitivity of both EARRs and IEARRs is considerable in comparison to the fault free case, which will facilitate the fault detection stage.

*Scenario 3: Gradual fault case in the presence of disturbances and in the absence of measurement noises*

In this scenario, a gradual component fault (a fault leading to extra abnormal friction in the motor [36]) is injected to the system, which is shown in Fig.10.

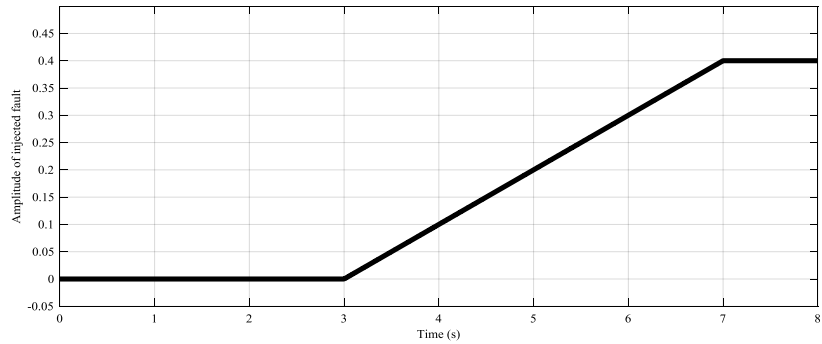

Fig.10 The injected gradual component fault

The obtained residuals are shown in Fig.11 and the residual evaluation functions are also given in Fig.12.

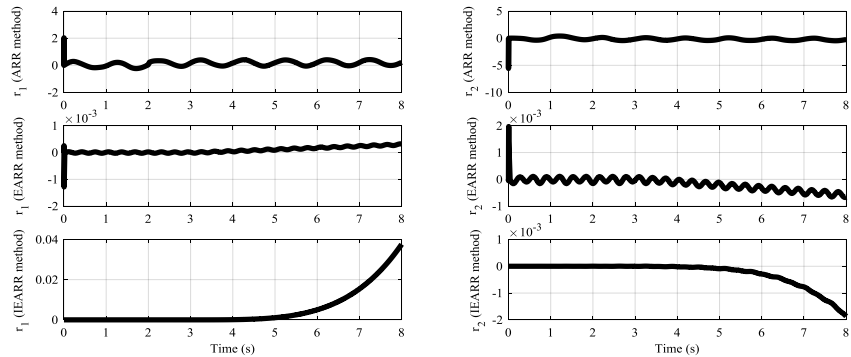

Fig.11: Residual for gradual component fault

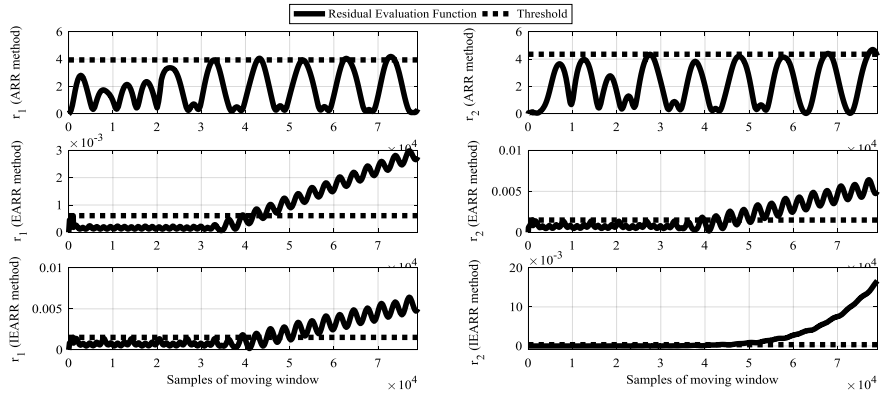

Fig.12: Residual evaluation function for gradual component fault

According to the figures, the residual evaluation functions of the proposed methods are efficiently exceeded from their predefined thresholds for the injected gradual fault as well. Hence, the fault can be detected rapidly by the proposed methods.

*Scenario 4: Fault free case in the presence of disturbances and measurement noises*

In this case, the performance of the proposed methods including EARRs and IEARRs are compared with the ARRs method in aspect of the robustness against the measurement noises. By considering the measurement noises as 2% of the output amplitude, the residuals are plotted as Fig.13 in fault free case.

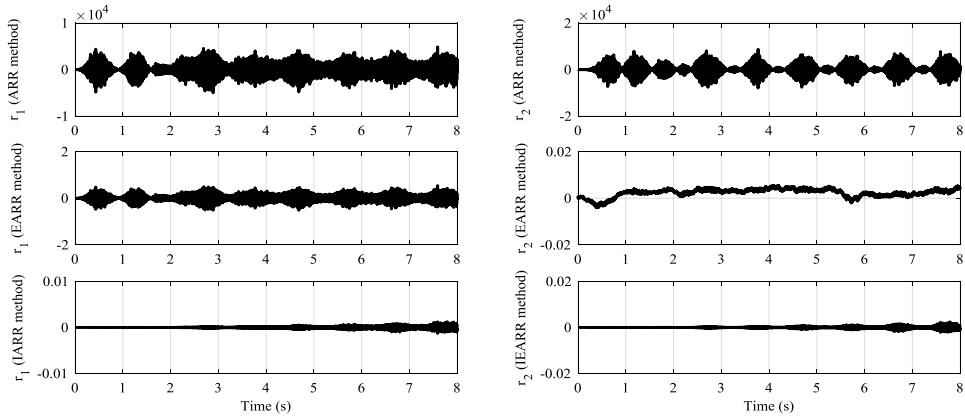

Fig.13: The residuals in fault free case in the presence of measurement noises

As it is clear from the figure, the residuals in IEARR method has better performance in the sense of robustness against measurement noises in comparison to the other methods. It is also worth noting that the ARR method has great sensitivity against measurement noises, which leads to large residuals compared to the first scenario.

*Scenario 5: Faulty case in the presence of disturbances and measurement noises*

The effects of the measurement noises on the residuals in faulty case are studied in this part of the paper. The residuals are given in Fig.14.

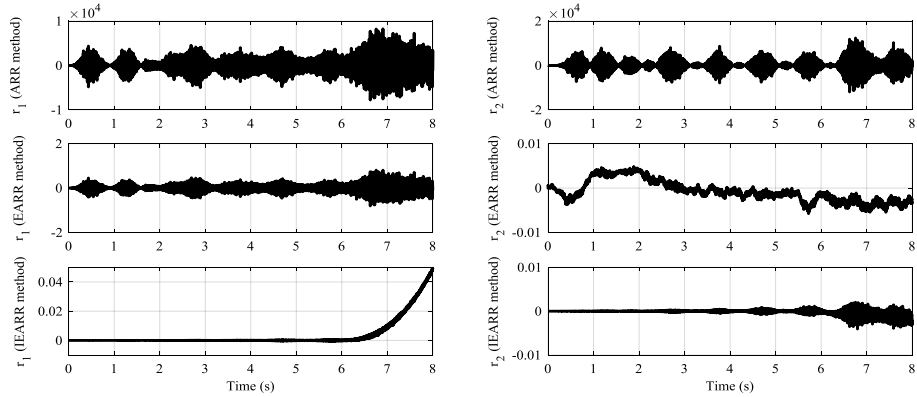

Fig. 14: The residuals in faulty case in the presence of measurement noises

As it can be observed from the figure, the IEARR method has great performance in this case as well. The injected fault can be detected efficiently in the presence of measurement noises by the integral form of the residuals.

According to the obtained figures in the considered scenarios, it can be concluded that the proposed methods show the great performance in comparison to the conventional ARRs method. The effects of disturbances are attenuated in the proposed methods (EARRs and IEARRs) by using the robust observer. The effects of noises on the residuals are also reduced in the IEARRs method in comparison to the two other methods. It is worth noting that the noises must be filtered in the FD system, or may be reduced by the available stochastic observers such as Kalman filter, which not considered in this study. In another approach, the thresholds may be obtained by considering some stochastic specifications of the measurement noises in addition to disturbances effects.

## 5- Conclusion

In this paper, the robust fault detection scheme for the nonlinear Lipschitz systems has been studied by using the BG method combined with a robust observer. The robustness against disturbance for the FD system based on the BG method has been considered by the combined BG-observer based residuals. The linear form of the residuals has been obtained which has been provided the possibility of the integral form residuals definition, which is robust against measurement noises. Two robust methods including EARRs and IEARRs have been presented for fault diagnosis of the Lipschitz nonlinear systems based on the BG method. The simulation results for a single link manipulator have been shown the great performance of the proposed methods in the presence of disturbances and measurement noises compared to the conventional ARR method. The obtained results have been proved the efficiency of the proposed method by simultaneously considering the disturbances, noises and some parametric uncertainties in the Lipschitz nonlinear systems.

## 6- Conflict of Interest

# 7- References